\documentclass[11pt, oneside]{article}   	% use "amsart" instead of "article" for AMSLaTeX format
\usepackage{geometry}                		% See geometry.pdf to learn the layout options. There are lots.
\usepackage{graphicx}				% Use pdf, png, jpg, or eps§ with pdflatex; use eps in DVI mode
\usepackage{amssymb}

\newtheorem{lemma}{Lemma}

\newtheorem{assumption}{Assumption}

\begin{document}

\title{New algorithm for the discrete logarithm problem on elliptic curves}

\author{Igor Semaev\\Department of Informatics\\ University of Bergen, Norway\\e-mail: igor@ii.uib.no\\ phone: (+47)55584279\\ fax: (+47)55584199}
%\date{}							% Activate to display a given date or no date

\maketitle
\begin{abstract}
A new algorithms for computing discrete logarithms on elliptic curves defined over finite fields is suggested. It is based on a new method to find zeroes of summation polynomials. In  binary elliptic curves one is to solve a cubic system of Boolean equations. Under a first fall degree assumption 
the regularity degree of the system is at most $4$. Extensive experimental data which supports the assumption is provided. An heuristic analysis suggests a new asymptotical complexity bound  $2^{c\sqrt{n\ln n}}, c\approx 1.69$ for computing  discrete logarithms on an elliptic curve over a field of size $2^n$. For several  binary elliptic curves recommended by FIPS the new method  performs better   than Pollard's.

\end{abstract}
%\section{}
%\subsection{}
\section{Introduction} Let $E$ be an elliptic curve defined over a finite field $F_q$ with $q$ elements. The discrete logarithm problem is given $P,Q\in E(F_q)$ compute an integer number $z$ such that $Q=zP$ in the group $E(F_q)$. That problem was introduced in \cite{M85,K87}. A number of information security standards are now based on the hardness of the problem, see \cite{FIPS} for instance. Two cases are of most importance: $q=p$ is a large prime number and $q=2^n$, where $n$ is prime. For super-singular and anomalous elliptic curves the discrete logarithm problem is easy, that was independently discovered by several authors, see \cite{MOV,Sem93,FR94} and \cite{Sem98,SA98,Smart99}. The more general are Pollard's methods \cite{Pollard}. They are applicable  to compute  discrete logarithms in any finite group. In elliptic curve case  the time complexity is proportional to $q^{1/2}$ field operations and the memory requirement is negligible. The method was  improved in \cite{ WZ98,GLV00,ET01} though   the asymptotical complexity bound remained.  In \cite{OW99}  a method for efficient parallelization of Pollard type algorithms  was provided.
 
Summation polynomials for elliptic curves were introduced in \cite{Sem04}. It was there suggested to construct an index calculus type algorithm for the elliptic curve discrete logarithm problem by decomposing points via computing zeroes of these polynomials. In \cite{G09,Diem11,JV12,FPPR12}   Gr\"{o}bner basis type algorithms were applied  for computing zeroes of summations polynomials or their generalisations over extension finite fields. For curves over some such  fields  the problems was proved to be sub-exponential, \cite{Diem11}. However no improvement for elliptic curves over  prime fields or binary fields of prime extension degree was achieved in those papers.

Based on  observations in \cite{FPPR12}, it was shown in   \cite{PQ12}  that under a first fall degree assumption for  Boolean equation systems coming from summation polynomials, the time complexity for elliptic curves over $F_{2^n}$ is sub-exponential and proportional to $2^{cn^{2/3}\ln n}$, where $c=2\omega/3$, and $2.376\leq\omega\leq 3$ is the linear algebra constant. It was there found that for $n>2000$ the method is better than Pollard's. The assumption was supported by experiments with computer algebra package MAGMA in \cite{PQ12,ST13}.

In this work we suggest computing zeroes of summation polynomials by solving a much simpler system of Boolean equations. The system incorporates more variables than previously but has algebraic degree only $3$. The first fall degree is proved to be $4$. Then a first fall degree assumption says the regularity degree $d_{F4}$ of the Gr\"{o}bner basis algorithm $F4$  is at most $4$ as well. The assumption was endorsed by numerous experiments with MAGMA. The new method overcomes strikingly what was achieved in the experiments of \cite{PQ12,ST13}.

   The time and memory complexity of computing summation polynomial zeroes under the assumption is polynomial in $n$.
The overall time complexity of computing  discrete logarithms on elliptic curves over $F_{2^n}$ becomes proportional to $$2^{c\sqrt{n\ln n}},$$ where $c=\frac{2}{(2\ln\, 2)^{1/2}}\approx 1.69$. Our analysis suggests a number of FIPS binary elliptic curves in \cite{FIPS} are theoretically broken as the new method starts to perform better than Pollard's for $n> 310$. The estimate is obviously extendable to  elliptic curves over $F_{p^n}$ for fixed $p>2$ and growing $n$, by using first fall degree bounds from \cite{HPS}. The time complexity is then $p^{c\sqrt{n\ln n}},$ where $c=\frac{2}{(2\ln\, p)^{1/2}}$.

\section{Summation polynomials and index calculus on elliptic curves}\label{S_polynomials}
Let $E$ be an elliptic curve over a field $K$ in Weierstrass form
\begin{equation}
Y^2+a_1XY+a_3Y=X^3+a_2X^2+a_4X+a_6,
\end{equation}
 For an integer $m\geq 2$ the $m$-th summation polynomial is the  polynomial $S_m$ in $m$ variables defined by the following property. Let $x_1,x_2,\ldots,x_m$ be any elements from $\bar{K}$, the algebraic closure of $K$, then $S_m(x_1,x_2,\ldots,x_m)=0$ if and only if there exist $y_1,y_2,\ldots,y_m\in \bar{K}$ such that the points $(x_i,y_i)$ are on $E$ and $$(x_1,y_1)+(x_2,y_2)+\ldots +(x_n,y_n)=\infty$$ in the group $E(\bar{K})$, see \cite{Sem04}. It is enough to find $S_3(x_1,x_2,x_3)$ then 
 for $m\geq 4$ in any case
\begin{equation}\label{resultant}
S_m(x_1,\ldots,x_m)=\hbox{Res}_X(S_{m-r}(x_1,\ldots,x_{m-r-1},X),S_{r+2}(x_{m-r},\ldots,x_{m},X)
\end{equation}
where $1\leq r\leq m-3$. The polynomial $S_m$ is symmetric for $m\geq 3$ and has degree $2^{m-2}$ in each its variable. $S_3$ was explicitly constructed in \cite{Sem04} for characteristic $\geq 5$ and characteristic $2$, the latter in case of a so called Koblitz curve. In characteristic $\geq5$ we can assume $a_1=a_3=a_2=0$ and denote $A=a_4,B=a_6$. So  
$$S_3(x_1,x_2,x_3)=(x_1-x_2)^2x_3^2-
2[(x_1+x_2)(x_1x_2+A)+2B]x_3+
(x_1x_2-A)^2-4B(x_1+x_2).$$
We are mostly concern with characteristic $2$ case and the curves recommended by \cite{FIPS}. So we can assume $a_1=1,a_3=0,a_4=0$ and denote $B=a_6$. Then $$S_3(x_1,x_2,x_3)=(x_1x_2+x_1x_3+x_2x_3)^2+x_1x_2x_3+B,$$
see \cite{Sem04,Diem11}.

 It was suggested in \cite{Sem04} to construct an index calculus type algorithm for the discrete logarithm problem in $E(F_q)$ via finding zeroes of summation polynomials. For random integer $u,v$ compute an affine point $R=uP+vQ=(R_X,R_Y)$. Then solve the equation
 \begin{equation}\label{OldEquations}
S_{m+1}(x_1,\ldots,x_m,R_X)=0.
\end{equation} for $x_i\in V$, where $V$ is a subset of $F_q$. Each solution provides with a linear relation(decomposition) which incorporates $R$ and at most $m$ point from a relatively small set of points in $E(F_q)$, whose $X$-coordinate belongs to $V$ and possibly an order $2$ point in $E(F_q)$.
 Then linear algebra step finds the unknown logarithm.
 Two cases were considered in \cite{Sem04}. First, $q$ is a prime number, then $V$ is a set of residues modulo $q$ bounded by $q^{1/n+\delta}$ for a small $\delta$. Second,  $q=2^n$, and $f(X)$ be an irreducible polynomial of degree $n$ over $F_2$, and
$F_{2^n}=F_2[X]/(f(X))$. Then $V$ is a set of all degree $<n/m+\delta$ polynomials modulo $f(X)$. However no algorithm to find the zeros in $V$ of summation polynomials  was suggested in  \cite{Sem04}. In the next Section \ref{Algorithm} we suggest producing the decomposition 
 by solving a different equation  system.  The new system is essentially equivalent to (\ref{OldEquations}). In case $q=p^n$, where $n$ is large, after reducing  the equations over $F_q$ to coordinate equations over $F_p$(Weil descent) the solution method is a Gr\"{o}bner basis algorithm. In Section \ref{Analysis} for   $p=2$  we show under a first fall degree assumption that the complexity of  a Gr\"{o}bner basis algorithm on such instances is polynomial. The assumption was proved correct in numerous experiments with computer algebra package MAGMA. A similar assumption looks  correct for odd $p$ too but the computations are tedious already for $p=5$. 

 \section{New algorithm}\label{Algorithm}
 Let $P$ be a point of order $r$ in the group $E(F_{q})$, where $E$ is an elliptic curve defined over $F_q$. 
 Then $Q=zP$ belongs to the subgroup generated by $P$. The discrete logarithm problem is given $Q$ and $P$, find $z\, \hbox{mod}\, r$.
In this section an algorithm for computing  $z$  is described.
\begin{enumerate}
\item Define parameter $m$ and a subset $V$ of $F_{q}$ of size around $q^{1/m}$.
\item For random integer $u,v$ compute $R=uP+vQ$. If $R=\infty$, then compute $z$ from the equation $bz+a\equiv 0\, \hbox{mod}\, r$. Otherwise, $R$ has affine coordinates $(R_X,R_Y)$. If $R_X=x_1\in V$,  then we have a relation (\ref{relation}) for $t=1$. Otherwise
\item for $t=2,\ldots,m$ try to compute $x_1,\ldots,x_t\in V$ and $u_1,\ldots,u_{t-2}\in F_{q}$ until the first system of  the following $t-1$ equations is satisfied
\begin{equation}\label{NewEquations}
\begin{array}{lccc}
S_{3}(u_1,x_1,x_2)&=&0,& \\
S_{3}(u_{i},u_{i+1},x_{i+2})&=&0,&1\leq i\leq t-3\\
S_{3}(u_{t-2},x_t,R_X)&=&0.&
\end{array}
\end{equation}
For $t=2$ the system consists of only one equation $S_3(x_1,x_2,R_X)=0$.
If non of the systems is satisfied repeat the step with a new $R$.
The solutions to   (\ref{NewEquations}) are  solutions to 
\begin{equation}\label{OldEquations_1}
S_{t+1}(x_1,x_2,\ldots,x_t,R_X)=0.
\end{equation}
 The reverse statement is true as well
if the  systems (\ref{NewEquations}) with lower $t$ are not satisfiable, 
see Lemma \ref{equiv}. In practical terms it is enough to solve only one system (\ref{NewEquations}) for $t=m$.   The experiments in characteristic $2$ presented below demonstrate that for $t<m$ the solving running time with a  Gr\"{o}bner basis algorithm drops dramatically and  the probability of solving is relatively  lower.  So it may be more efficient to solve a lot of the systems with $t<m$ for different $R$ instead of one system for $t=m$ with one $R$. One can probably win in efficiency and lose in  probability. Though the trade off may be positive, we won't pursue  this approach in the present work as this does not affect the asymptotical running time estimates.
 
Compute $y_1,\ldots,y_{t}\in F_{q^2}$ such that
\begin{equation}\label{relation}
(x_1,y_1)+(x_2,y_2)+\ldots+(x_t,y_t)+uP+vQ=\infty.
\end{equation}
If there are $y_i\in F_{q^2}\setminus F_q$, then  the sum of all points $(x_i,y_i)$ in (\ref{relation}), where $y_i\in F_{q^2}\setminus F_q$, is a point  in $E(F_q)$ of  order exactly $2$, see Lemma \ref{equiv}. So that is a useful relation anyway.
At most $|V|$ relations (\ref{relation}) are necessary on the average.

\item Solve the linear equations (\ref{relation}) and get  $z\, \hbox{mod}\, r$. 
\end{enumerate}

\section{Analysis}\label{Analysis}
 We will show in Lemma \ref{equiv}  that the system (\ref{NewEquations})  is essentially equivalent to (\ref{OldEquations_1}).
  Despite the number of variables in  (\ref{NewEquations}) is significantly larger, each  equation on its own is much simpler. In particular, the algebraic degree of equations (\ref{NewEquations}) in each of the variables is only $2$ in contrast to $2^{t-1}$ for   (\ref{OldEquations_1}).
  \subsection{Lemmas}\label{lemmas}  
 \begin{lemma}\label{equiv_1}Let the elliptic curve $E$ be defined over a field $F_q$.
 Let $x_1,\ldots,x_t\in V$ be a solution to \textup{(\ref{OldEquations_1})}. Then there exist $y_1,\ldots,y_t\in F_{q^2}$ such that
\begin{equation}\label{relation_1}
(x_1,y_1)+(x_2,y_2)+\ldots+(x_t,y_t)+R=\infty.
\end{equation}
\end{lemma} 
 
 \begin{lemma}\label{equiv}
Let $R_X\notin V$. Assume the equations $$S_{i+1}(x_1,\ldots,x_{i},R_X)=0,x_1,\ldots,x_i\in V$$ are not satisfiable for $2\leq i<t$ and  $x_1,\ldots,x_t\in V$ is a solution to $S_{t+1}(x_1,\ldots,x_{t},R_X)=0$. Then

\begin{enumerate}

\item in \textup{(\ref{relation_1})}  
assume   $y_1,\ldots,y_s\in F_{q^2}\setminus F_q$ and $y_{s+1},\ldots,y_t\in  F_q$.  Then
$$H=(x_1,y_1)+\ldots+(x_s,y_s)$$
is a point in $E(F_q)$ of order exactly $2$. So  $s=0$ or $s\geq 2$.

\item   There exist $u_1,\ldots,u_{t-2}\in F_q$ such that 
\begin{equation}\label{NewEquations_1}
\begin{array}{lccc}
S_{3}(u_1,x_1,x_2)&=&0,& \\
S_{3}(u_{i},u_{i+1},x_{i+2})&=&0,&1\leq i\leq t-3\\
S_{3}(u_{t-2},x_t,R_X)&=&0.&
\end{array}
\end{equation}
\end{enumerate}
    
\end{lemma}
\emph{Proof} Let's prove the first statement of the lemma. Assume $s>0$ and let
$$G=(x_{s+1},y_{s+1})+\ldots+(x_t,y_t)\in E(F_q).$$
Then $H=-R-G\in E(F_q)$ as well. Let $\phi$ be a non-trivial automorphism of $F_{q^2}$ over $F_q$, then 
$$ \phi(H)+H=\infty,\quad \phi(H)=H,$$
and so $2H=\infty$. If $H=\infty$, then $G+R=\infty$ and so $S_{t-s+1}(x_{s+1},\ldots,x_{t},R_X)=0$. That contradicts the assumption. Therefore $H$ is a point in $E(F_q)$ of order exactly $2$ and $s\geq 2$.

Let's prove the second statement. Assume   $y_1,\ldots,y_s\in F_{q^2}\setminus F_q$ and $y_{s+1},\ldots,y_t\in  F_q$, where $s=0$ or $s\geq 2$. There are points $P_1,\ldots,P_{t-2}$ such that
\begin{equation}\label{NewEquations_equiv}
\begin{array}{lllccr}
(x_1,y_1)&+(x_2,y_2)&+P_1&=&\infty,&\\
P_i&+(x_{i+2},y_{i+2})&+P_{i+1}&=&\infty,&1\leq i\leq t-3\\
P_{t-2}&+(x_t,y_t)&+R&=&\infty.&\\
\end{array}
\end{equation}
By the lemma assumption and the previous statement $P_1,\ldots,P_{t-2}\ne\infty$. So $P_i=(u_i,v_i)$, where $u_i\in F_q$. Therefore  (\ref{NewEquations_equiv}) implies (\ref{NewEquations_1}). The lemma is proved.

A variation of the first  statement of Lemma \ref{equiv} has already appeared in \cite{Sem04}.

\subsection{General discussion on complexity}  The complexity of solving a linear system of equations (\ref{relation}) is taken $O(|V|^{\omega'})$, where   $\omega'=2$ as the system is very sparse for any finite field $F_q$, see \cite{Wied}.

 It is not quite clear how the system (\ref{NewEquations} may be resolved in case $q$ is a large prime number. However, 
 for $q=p^n$, where  $n$ is large, a  Gr\"{o}bner basis algorithm is applicable. The approach was already
  used in \cite{G09,Diem11} for solving (\ref{OldEquations}) after it was reduced to a system of $n$ multivariate polynomial equations in about $n$ variables over $F_p$ by so called Weil descent, where $V$ may be taken any vector space of dimension $k=\lceil n/m\rceil$ over $F_p$.  The problem of generating such a system and keeping it in computer memory before solving is difficult by itself for $m\geq 4$ and the difficulties increase rapidly for larger $m$. In \cite{PQ12} it was shown that the complexity of solving (\ref{OldEquations}) is sub-exponential in $n$ under a first fall degree assumption, see Section \ref{ffd} below.  That assumption was supporter by a number of experiments in \cite{PQ12,ST13}, where the parameter  $m$ was taken at most $3$. 
 
 In this paper we suggest using a  Gr\"{o}bner basis algorithm
  to solve (\ref{NewEquations}) rather than (\ref{OldEquations}). The system (\ref{NewEquations}) for $t=m$   is equivalent to a system of $(m-1)n$ multivariate equations in  $(m-2)n+km\approx (m-1)n$ variables in $F_p$. Under a first fall degree assumption, see Section \ref{ffd} and Assumption \ref{Assumption}, we show its complexity is $O[(n(m-1))^{4\omega}]$, where $2.376\leq \omega\leq 3$, that is  
 polynomial in $n$.  The assumption was proved correct in numerous experiments with MAGMA, see Section \ref{Experiments}. We were able to solve (\ref{NewEquations}) for $t=m$ and therefore  (\ref{OldEquations}) for $m$ as $5,6$ and some $n$   on a common computer. For  $n,m$ as in \cite{PQ12,ST13}, the solution is up to $50$ times faster and takes up to $10$ times less memory  in comparison with \cite{PQ12,ST13}. Similar to \cite{PQ12}, one can   take the advantage of a block structure of the Boolean system resulted from (\ref{NewEquations}), though that does not affect the asymptotical estimates. 
  By extrapolating running time estimates we find that four binary curves recommended by FIPS PUB 186-4 \cite{FIPS} for $n=409,571$ become theoretically broken as the new method is faster than Pollard's for $n>310$, see Section \ref{Asymptotical}. If  the block structure of the system is not exploited and we extrapolate the complexity of the default Gr\"{o}bner basis algorithm F4, then only two FIPS curves for $n=571$ are broken.
  
     In practical terms 
an additional  effort is required in order to accelerate the decomposition stage by solving  (\ref{NewEquations}) and to break the rest of the binary curves  in \cite{FIPS}. As the collecting stage still significantly dominates the method running time, see Table \ref{Time_Extrapolation}, one can use almost unbounded parallelisation to get more efficiency.
 In asymptotical analysis the complexity of generating summation polynomials and computing their zeros to get point decomposition may be neglected as it is polynomial. 
That significantly improves the asymptotical complexity bound in \cite{PQ12}, see Section \ref{Asymptotical}.
   \subsection{Success probability} \label{Probability}
    We  estimate 
    the probability    that
  $$S_{t+1}(x_1,\ldots,x_t,R_X)=0,x_1,\ldots,x_t \in V, $$ where $2\leq t\leq m$, is satisfiable. We adopt the following model. For random $z$ the mapping $ x_1,\ldots,x_t \rightarrow 
S_{t+1}(x_1,\ldots,x_t,z)$ is a symmetric random mapping from $V^t$ to $F_{q}$. Let $K$ be the number of classes of tuples $(x_1,\ldots,x_t)$  under permuting the entries. Then $K\approx \frac{|V|^t}{t!}$. The probability of a solution is the probability $P(q,m,t,|V|)$ that the mapping hits $0\in F_q$ at least once. So
\begin{eqnarray}
P(q,m,t,|V|)&=&1-(1-1/q)^{K}\nonumber\\
&\approx& 1-(1-1/q)^{\frac{|V|^t}{t!}}\approx 1-e^{-\frac{|V|^t}{q\,t!}}
\end{eqnarray}
If $\frac{|V|^t}{q\,t!}=o(1)$, then $P(q,m,t,|V|)\approx \frac{|V|^t}{q\,t!}$ as in \cite{Diem11,PQ12}. 
 It is obvious the probability of solving at least one of the first $t-1$  systems (\ref{NewEquations}) is at least  $ P(q,m,t,|V|)$. 
On the other hand, the latter is larger than the probability of solving (\ref{NewEquations}). Therefore, we can assume that the probability of solving (\ref{NewEquations}) is approximately $ P(q,m,t,|V|)$. 
 
  In case $q=p^n$ we denote the probability $P(q,m,t,|V|)$ by $P(n,m,t,k)$, where $|V|=p^k$ and $p$ should be clear from the context. 
  
 \subsection{Solving polynomial equations and first fall degree assumption }\label{ffd}
 Let 
 \begin{eqnarray}\label{system}
f_1(x_1,\ldots,x_n)=0,\nonumber\\
f_2(x_1,\ldots,x_n)=0,\nonumber\\
\ldots\\
f_m(x_1,\ldots,x_n)=0,\nonumber
\end{eqnarray}
be a system of polynomial equations over a field $K$.  The system (\ref{system}) may be solved by first finding a Gr\"{o}bner basis $g_1,g_2,\ldots,g_s$ for the ideal generated by polynomials $f_1,f_2,\ldots,f_m$. If the ground field $K=F_q$ is a finite field of $q$ elements and we want the solutions with entries in $F_q$, then the basis is computed for the ideal generated by
$$f_1,f_2,\ldots,f_m,\,x_1^q-x_1,\ldots,x_n^q-x_n.$$
 The solutions to $g_i(x_1,\ldots,x_n)=0, (1\leq i\leq s)$ are  solutions to (\ref{system}) and they are relatively easy to find due to the properties of the Gr\"{o}bner basis. Several algorithms were designed to construct a Gr\"{o}bner basis. Let $\deg g$ denote the total degree of the polynomial $g=g(x_1,\ldots,x_n)$. The first algorithm  \cite{bB76} was based on reducing pairwise combinations(S-polynomials) of  the polynomials from the current basis and augmenting the current basis with their remainders. Equivalently \cite{dL83}, one can triangulate a Macaulay matrix $M_d$ whose rows are coefficients of the polynomials $m_if_j$, where $m_i$ are monomials and $\deg(m_i)+\deg(f_i)\leq d$ for a parameter $d$. That produces a Gr\"{o}bner basis for some large enough $d=d_0$. The matrix incorporates at most ${n+d-1\choose{d}}< n^d$ columns.  So the complexity is $O(n^{\omega d_{0}})$ of the ground field operations, where $2.376\leq \omega\leq 3$ is the linear algebra constant.

 Also  one may solve a system of linear equations which comes from $m_if_j=0,\deg(m_i)+\deg(f_i)\leq d$ after linearisation  to get the solutions to (\ref{system}) without computing a Gr\"{o}bner basis. The method is called extended linearisation(XL).
 The matrix of the system is essentially $M_d$. For large enough $d=d_{1}$ the rank of the matrix is close to the number of variables after linearisation \cite{YCC04}.   
 The complexity is  $O(n^{\omega' d_{1}})$ of the ground field operations, where $2\leq \omega'\leq 3$ is a linear algebra constant, which depends on the sparsity of the matrix. It may be that $\omega'=2$ for a very sparse matrix  in case of solving by extended linearisation.
 
Numerous experiments with solving the equations (\ref{NewEquations}) by computer algebra package MAGMA were done in this work. MAGMA implements an efficient Gr\"{o}bner basis algorithm F4\cite{F99}. The algorithm  successively constructs Macaulay type matrices of  increasing sizes, compute row echelon forms of them, produce some new polynomials and use them in the next step of the construction as well. At some point no new polynomials are generated. Then the current set of  polynomials is a Gr\"{o}bner basis. The complexity is characterised by  $d_{F4}$, the maximal total degree of the polynomials occurring before a Gr\"{o}bner basis is computed.  The overall complexity  is the sum of the complexities of some steps, where 
  the largest step complexity is bounded by 
    \begin{equation}\label{complexity}
O(n^{\omega\, d_{F4}}).
\end{equation}
 We assume the complexity of the computation is determined by the complexity of the largest step. In the experiments in Section \ref{Experiments} the ratio between the overall running time and the largest step running time was bounded by   $\approx 3$ for the number of variables  $n\approx 50$. So in the asymptotical analysis below we accept (\ref{complexity}) as the complexity of F4. 
   Another Gr\"{o}bner basis algorithm F5\cite{F02} with the maximal total degree $d_{F5}$ has the complexity $O(n^{\omega d_{F5}})$, see \cite{BFS03}. It was implicitly assumed in \cite{PQ12,ST13} that $d_{F5}=d_{F4}$.

 We will use the following definition found in \cite{PQ12}. The first fall degree for (\ref{system}) is the smallest total degree $d_{ff}$ such that there exist polynomials $g_i=g_i(x_1,\ldots,x_n),(1\leq i\leq m)$ with $$max_i(\deg g_i+\deg f_i)=d_{ff},\quad \deg \sum_ig_if_i<d_{ff}$$ and $\sum_ig_if_i\ne 0$. A first fall degree assumption says $d_{F4}\leq d_{ff}$ and that is a basis for asymptotical complexity estimates in  \cite{PQ12}. Although  not generally correct, the assumption
  appears correct for the polynomial systems coming from (\ref{OldEquations}) 
   and was supported by  extensive experiments for relatively small parameters in  \cite{PQ12,ST13}. This is very likely correct for (\ref{NewEquations}) as well, see sections below.

\subsection{Characteristic 2} Let  $E$ be determined by
$$Y^2+XY=X^3+AX^2+B,$$
$A,B\in F_{2^n}$. Therefore, 
\begin{equation}\label{S_polynomial}
S_3(x_1,x_2,x_3)=(x_1x_2+x_1x_3+x_2x_3)^2+x_1x_2x_3+B,
\end{equation}
see Section \ref{S_polynomials}. Let $f(x)$ be an irreducible polynomial of degree $n$ over $F_2$ and $\alpha$  its root in $F_{2^n}$. Then $1,\alpha,\ldots,\alpha^{n-1}$ is a basis of $F_{2^n}$ over $F_2$. Elements of $F_{2^n}$ are represented as  polynomials in $\alpha$ of degree at most $n-1$. Let  $V$ be a set of all polynomials in $\alpha$ of degree $<k=\lceil n/m\rceil$.  Obviously, that is a vector space over $F_2$ of dimension $k$. Following \cite{Diem11}, one can define $V$ as any subspace of $F_{2^n}$ of  dimension $k$. However it seems that using the subspace of low degree polynomials significantly reduces the time and space complexity  in comparison with a randomly generated subspace and is therefore preferable. We attribute the phenomena to the fact that the set of polynomials to compute a Gr\"{o}bner basis is simpler in the former case.  

According to  \cite{G09,Diem11}  the equation (\ref{OldEquations}),
where $x_i\in V$, is reducible, by taking coordinates(so called Weil descent), to a system of $n$ Boolean equations in $mk$ variables. A Gr\"{o}bner basis algorithm is applicable to find its solutions. The maximal total degree of the Boolean equations  is at most $m(m-1)$. Also it was observed in \cite{PQ12} and proved in \cite{HPS} that the first fall degree of the Boolean equations coming from
$$S_{m+1}(x_1,\ldots,x_m,R_X)=0$$
is at most $m^2+1$.

 We consider the case $m=2$ in more detail now. First we take the polynomial (\ref{S_polynomial}), where all $x_1,x_2,x_3$ are variables in $V$ or $F_{2^n}$. Following an idea in \cite{PQ12} it is easy to prove  that the first fall degree is $4$ in this case. Really, coordinate Boolean functions which represent $S_3(x_1,x_2,x_3)$ are of total degree $3$. We denote that  fact
  $$\deg_{F_2} S_3(x_1,x_2,x_3)=3.$$ 
  However $$\deg_{F_2} x_1S_3(x_1,x_2,x_3)=3$$ again because
\begin{eqnarray}
x_1S_3(x_1,x_2,x_3)&=&x_1[(x_1x_2+x_1x_3+x_2x_3)^2+x_1x_2x_3+B]\nonumber\\
&= &x_1^3x_2^2+x_1^3x_3^2+x_1x_2^2x_3^2+x_1^2x_2x_3+Bx_1\nonumber
\end{eqnarray}
despite $\deg_{F_2}x_1+\deg_{F_2} S_3(x_1,x_2,x_3)=4$. This argument does not work for $S_3(x_1,x_2,z)$, where $z$ is a constant from $F_{2^n}$. We have 
\begin{eqnarray}
\deg_{F_2} S_3(x_1,x_2,z)=2,\nonumber\\
\deg_{F_2} x_1S_3(x_1,x_2,z)=3,\nonumber
\end{eqnarray}
and $\deg_{F_2}x_1+\deg_{F_2} S_3(x_1,x_2,z)=3$. The first fall degree for such polynomials was bounded by $5$ in \cite{HPS}. The experiments  show it is always $4$ again.
Anyway at least $t-2$ of the    equations in (\ref{NewEquations}) have the first fall degree $4$. So
we come up with the following assumption.
\begin{assumption}\label{Assumption}
Let $q=2^n$ and $2\leq m< n$, $k=\lceil \frac{n}{m}\rceil$. Also let $V$ be a subspace of dimension $k$ in $F_{2^n}$. Then  $d_{F4}\leq 4$  for a Boolean equation system equivalent to \textup{(\ref{NewEquations})} for any $2\leq t\leq m$.
\end{assumption}

\subsubsection{Experiments}\label{Experiments}
In this section we check   Assumption \ref{Assumption} by experiments with  MAGMA. The package implements the  Gr\"{o}bner basis type algorithm F4 due to Faug\`{e}re \cite{F99}. We run the algorithm  to construct solutions for the system of Boolean equations  resulted from (\ref{NewEquations}). The system consists of $n(t-1)$ coordinate equations in $n(t-2)+kt$ variables and $n(t-2)+kt$ field equations are added. To simplify computations the $X$-coordinate 
 $R_X$ of a random $R$ was substituted by a random element $z$ from $F_{2^n}$.  We take the  parameters $n,t\leq m,k=\lceil n/m\rceil$ from a range of values and solve the system  for   $100$ random $z$.  

The results, where $B=1$ in (\ref{S_polynomial}), are presented in Table \ref{Main_1}. The results, where $B$ is a randomly generated element from $F_{2^n}$, are presented in Table \ref{Main_2} . In the columns of the tables the following parameters are shown: $n, m,t, k=\lceil \frac{n}{m}\rceil$, the experimental success probability, theoretical success probability $P(n,m,t,k)$, maximal degree $d_{F4}$ of the polynomials generated by F4 before a Gr\"{o}bner basis is computed, average time in seconds for solving one system  and overall amount of memory in MB used for solving $100$ systems (\ref{NewEquations}).  A computer with 2.6GHz Intel Core i7 processor and 16GB 1600MHZ DDR3 of memory was used. The most important of all the parameters is $d_{F4}$. We use the verbosity implemented in MAGMA for Faug\`{e}re's F4. The computation by F4 is split into a number of steps, where "step degree" is the maximal total degree of the polynomials for which a row echelon form is computed. The parameter is available for every step of the algorithm. If the ideal generated by the polynomials is unit, then "step degree" was always bounded by $4$. If not, that is there is a solution, then  "step degree" was bounded  by $4$ for all the steps before the basis is computed. They were followed  by at most three more steps, where "step degree" was $5,6,7$ with the message  "No pairs to reduce". At this point the computation stops.
\begin{table}[htdp]
\caption{Max. degree of polynomials  by MAGMA and other parameters, $B=1$.}
\begin{center}
\begin{tabular}{|l|c|c|c|r|c|r|r|}
\hline
$n$&$t=m$&$k=\lceil n/m\rceil$&exp. prob.&$P(n,m,t,k)$ & $d_{F4}$&av. sec.&MB\\\hline
12&6&2&0.00&0.0013&4&2.30&257.8\\\hline
13&4&4&0.41&0.2834&4&81.64&739.8\\\hline
13&5&3&0.05&0.0327&4&84.65&1597.8\\\hline
14&4&4&0.10&0.1535&4&79.57&879.7\\\hline
14&5&3&0.02&0.0165&4&23.47&960.2\\\hline
15&4&4&0.11&0.0799&4&136.70&1457.3\\\hline
15&5&3&0.05&0.0082&4&300.48&3286.9\\\hline
16&4&4&0.09&0.0408&4&175.72&1657.7\\\hline
17&3&6*&0.32&0.2834&4&27.08&378.1\\\hline
17&3&6&0.30&0.2834&4&11.41&364.1\\\hline
17*&3&6&0.36&0.2834&4&12.09&355.1\\\hline
17*&3&6*&0.25&0.2834&4&34.32&693.2\\\hline
\end{tabular}
\end{center}
\label{Main_1}
\end{table}%
To fill the tables 2300 Boolean systems each of total degree $3$ coming from (\ref{NewEquations}), where $t=m$, were solved. For all of them the maximal total degree attained by F4 to compute  a Gr\"{o}bner basis was exactly $4$. For $t<m$ the maximal total degree was   smaller or equal to $4$.
 We conclude that for all values of $n,m$ and $t\leq m$ in the tables  Assumption \ref{Assumption} was  correct for  randomly chosen $z\in F_{2^n}$. So the assumption is very likely to be correct for any values of $n,m,t\leq m$.
 
 The method significantly overcomes what was experimentally  achieved in \cite{PQ12,ST13}. For instance, $n=21, m=3,k=7$ the solution  in \cite{ST13}  of (\ref{OldEquations}) took 6910 seconds on the average with 27235 MB maximum memory used. With the new method  the solution of (\ref{NewEquations}) for $t=m$, and therefore (\ref{OldEquations}) as well, takes 133.5 seconds on the average and 2437.8 MB maximum memory on an inferior computer, see Table \ref{Main_2}.

\begin{table}[htdp]
\caption{Max. degree of polynomials  by MAGMA and other parameters, random $B$.}
\begin{center}
\begin{tabular}{|l|c|c|c|c|r|c|r|r|}
\hline
$n$&$m$& $t$&$k=\lceil n/m\rceil$&exp. prob.&$P(n,m,t,k)$ & $d_{F4}$&av. sec.&MB\\\hline
12&6&6&2&0.01&0.0013&4&2.52&289.9\\\hline\hline
13&4&4&4&0.31&0.2834&4&85.30&981.8\\\hline
13&5&5&3&0.09&0.0327&4&98.81&1633.0\\\hline\hline
14&4&4&4&0.16&0.1535&4&93.48&1056.7\\\hline
14&5&5&3&0.03&0.0165&4&41.15&1154.2\\\hline\hline
15&4&4&4&0.06&0.0799&4&102.85&1177.5\\\hline
15&4&3&4&0.02&0.0206&4&0.4765&64.1\\\hline
15&4&2&4&0.00&0.0038&4&0.0013&32.1\\\hline
15&5&5&3&0.03&0.0082&4&174.47&2635.4\\\hline
15&5&4&3&0.01&0.0051&4&12.95&424.9\\\hline
15&5&3&3&0.00&0.0026&4&0.0339&32.1\\\hline
15&5&2&3&0.00&0.0009&4&0.0006&32.1\\\hline\hline
16&4&4&4&0.04&0.0408&4&160.87&1145.2\\\hline
16&4&3&4&0.01&0.0103&4&0.4984&64.1\\\hline
16&4&2&4&0.00&0.0019&4&0.0014&32.1\\\hline\hline

17&3&3&6&0.21&0.2834&4&15.80&375.8\\\hline\hline
19&3&3&7&0.47&0.4865&4&137.32&1812.8\\\hline
19&3&2&7&0.01&0.0155&4&0.0092&32.1\\\hline\hline

21&3&3&7&0.12&0.1535&4&133.54&2437.8\\
\hline
21&3&2&7&0.01&0.0038&4&0.0095&32.1\\\hline
\end{tabular}
\end{center}
\label{Main_2}
\end{table}%

In the last 4 lines of Table \ref{Main_1} we also take into account the influence of the choice of generating polynomial for $F_{2^n}$ and  the vector space $V$. The line with $17*$ means a random irreducible polynomial $f(X)$ of degree $17$ for constructing $F_{2^{17}}$ was used in the computations. Otherwise a default generating polynomial of MAGMA or a sparse polynomial were used.  The line with $6*$ means a random subspace of dimension $6$ in $F_{2^{17}}$ was used in the computations. Otherwise a subspace of all degree $<6$ polynomials modulo $f(X)$ was used. We realise  the latter is preferable.  

To conclude the section we should mention that
 the maximal degree(regularity degree) generally exceeds $4$  when $k>\lceil \frac{n}{m}\rceil$ though the first fall degree is still $4$.
\subsubsection{Asymptotical Complexity}\label{Asymptotical}
 In this section an asymptotical complexity estimate for the  discrete logarithm problem in $E(F_{2^n})$ based on  Assumption  \ref{Assumption} is derived. The algorithm complexity is the sum of the complexity of two stages. First collecting a system of $\leq 2^k, k=\lceil n/m\rceil$ linear relations (\ref{relation}) and then solving them. The probability of producing one linear relation by solving at least one system  of multivariate Boolean equations  (\ref{NewEquations}) for $2\leq t\leq m$ is at least $P(n,m,m,k)\approx 2^{mk-n}/m!$, see Section \ref{Probability}. The complexity of solving by the Gr\"{o}bner basis algorithm F4 is 
$[n(m-1)]^{4\omega}$. The estimate  in \cite{PQ12} was based on using 
 a block structured Gr\"{o}bner basis algorithm, where the block size was $k$, rather than the standard F4. That reduced the asymptotical complexity of solving (\ref{OldEquations}).
  We think the same approach is applicable to solve the equations coming from (\ref{NewEquations}) as well, with the block size $n$. That reduces the complexity of finding the relation to  $n^{4\omega}$.  
  We remark that does not affect the asymptotical complexity of the present method anyway as the both estimates are polynomial.
Therefore the complexity of the first stage is
\begin{equation}\label{first_stage}
\frac{2^k n^{4\omega}}{P(n,m,m,k)}\approx \frac{m!}{2^{mk-n}}2^kn^{4\omega}
\end{equation}
operations, where $2.376\leq\omega\leq 3$ is the linear algebra constant. For $\omega=3$ that is at most $m!2^{\frac{n}{m}}n^{12}$.  The complexity of the second stage is 
\begin{equation}\label{second_stage}
2^{k\omega'},
\end{equation}
where $\omega'=2$ is the sparse linear algebra constant. One equates (\ref{first_stage}) and (\ref{second_stage}) to determine the optimal value  $m\approx \sqrt{\frac{(2\ln 2) n}{\ln n}}$ for large $n$. The overall complexity is
$$2^{k\omega'}\approx 2^{\frac{n\omega'}{m}}=
2^{c
\sqrt{n\ln n}},$$
where $c=\frac{2}{(2\ln 2)^{1/2}}$. 

We now compare the values of (\ref{first_stage}) and (\ref{second_stage}) for a range of $n\leq 571$ in Table \ref{Time_Extrapolation}. The first stage complexity  dominates. For each $n$ in the range one finds $m$, where the first stage complexity $m!2^{\frac{n}{m}}n^{12}$ is minimal. The table presents the values of $$n,\,2^{n/2},\,m,\,m!\,2^{\frac{n}{m}}\,n^{12},\, 2^{2n/m}.$$
 The new method starts performing better than  Pollard's for $n>310$. Therefore  four curves defined over $F_{2^n}$ for $n=409,571$ and recommended by FIPS PUB 186-4 \cite{FIPS} are theoretically broken. If only the default Gr\"{o}bner basis algorithm F4 is used with complexity 
 $[n(m-1)]^{4\omega}$, then two FIPS curves for $n=571$ are broken.

The first stage of the algorithm is easy to accomplish with several processors working in parallel. As the first stage complexity is dominating that significantly improves the running time of the method in practical terms. 
 
 \begin{table}[htdp]
\caption{Complexity estimates in characteristic 2.}
\begin{center}
\begin{tabular}{|c|c|c|c|c|}
\hline
n&$2^{n/2}$&m&$m!2^{\frac{n}{m}}n^{12}$&$2^{\frac{2n}{m}}$\\\hline
100&$1.12\times 10^{15}$&6&$7.49\times 10^{31}$&$1.08\times 10^{10}$\\\hline
150&$3.77\times 10^{22}$&7&$1.84\times 10^{36}$&$7.96\times 10^{12}$\\\hline
200&$1.26\times 10^{30}$&8&$5.54\times 10^{39}$&$1.12\times 10^{15}$\\\hline
250&$4.25\times 10^{37}$&9&$4.97\times 10^{42}$&$5.29\times 10^{16}$\\\hline
300&$1.42\times 10^{45}$&10&$2.07\times 10^{45}$&$1.15\times 10^{18}$\\\hline
310&$4.56\times 10^{46}$&10&$6.13\times 10^{45}$&$4.61\times 10^{18}$\\\hline
350&$4.78\times 10^{52}$&10&$4.21\times 10^{47}$&$1.18\times 10^{21}$\\\hline
400&$1.60\times 10^{60}$&11&$5.92\times 10^{49}$&$7.81\times 10^{21}$\\\hline
409&$3.63\times 10^{61}$&11&$1.36\times 10^{50}$&$2.43\times 10^{22}$\\\hline
450&$5.39\times 10^{67}$&11&$5.68\times 10^{51}$&$4.26\times 10^{24}$\\\hline
500&$1.80\times 10^{75}$&12&$4.08\times 10^{53}$&$1.21\times 10^{25}$\\\hline
571&$8.79\times 10^{85}$&12&$1.21\times 10^{56}$&$4.44\times 10^{28}$\\\hline

\end{tabular}
\end{center}
\label{Time_Extrapolation}
\end{table}%

\end{document}